\documentstyle[preprint,prb,aps,epsfig]{revtex}
\begin{document}
\headheight2.0cm
\headsep1.2cm
\baselineskip8mm
\tolerance=1500
\thispagestyle{empty}
\tightenlines
\begin{list}{}{\setlength{\leftmargin}{1mm}\baselineskip8mm}
\item
\begin{center}
\item INCOMPRESSIBLE STATES IN DOUBLE QUANTUM DOTS.
\item N. Barber\'an and J. Soto
\item Departament d'Estructura i Constituents de la Mat\`eria
\item Facultat de F\'\i sica, Universitat de Barcelona,
E-08028 Barcelona, Catalonia, Spain
\end{center}
\begin{abstract}

Incompressible (magic) states that result from
many-body effects in vertically coupled quantum dots
submitted to strong magnetic fields such that only the lowest Landau
level is relevant are studied within an exact
diagonalization calculation for $N=3,$ $5$ and $6$,
electrons.
We find that the sequences of total angular momentum $M$ for which
these incompressible states
exist depend on the interplay
between the inter-dot hopping parameter $\Delta_t$ and the inter-dot
distance
$d$. For $d$ of the order of the magnetic length and for all values
of
$\Delta_t$, we conclude
that, in contrast to previous claims, these incompressible states
appear at magic values of
$M$ which do not differ from those obtained for a
single dot, namely
$\,\,M\,=\,N(N-1)/2\,+\,j\,N\,\,$ where $j$ is a positive integer
number. For large inter-dot distance and simultaneously small
inter-dot hopping parameter, new sequences of magic values of $M$ are
observed. These new sequences can be easily understood in
terms of a transition regime towards  a system of two decoupled
single dots. However, important differences in
the nature of the incompressible ground states are found with respect
to those of a single dot.
\end{abstract}
\vskip2mm

\vfill
\hfill UB-ECM-PF 02/24
\end{list}
\eject

\subsection*{I. Introduction}

Much effort has been devoted to understand the magic
incompressible states (IS's) of two dimensional
electronic nanostructures. This is due to the fact
that they are closely related
to the states that determine
properties like superconductivity or the quantum Hall
effect (QHE) \cite{lau,lau1}, which are striking
examples of the non-trivial behavior that strongly interacting
electronic systems may display \cite{pra,yos}. Finite systems like
quantum dots (QD's)
provide simpler physical realizations of strongly interacting
electronic systems where different models can be tested.
When they are submitted to strong magnetic fields, the projection
of the system to the lowest Landau level (LLL) becomes a good
approximation which greatly
simplifies theoretical studies in general, and, in particular,
makes exact
diagonalization calculations feasible.
Much work has been done on single QD's in the LLL regime yielding
a reasonable understanding of the
nature of their IS's \cite{jac,cha}. The search of
IS
with well defined properties which may produce fractional QHE
experimentally observable, led to analyze double
layered systems \cite{cha1,mur,par,par1,ani,yan}.
Double quantum dots (DQD's) in
a vertical configuration submitted to strong magnetic fields provide
a finite system in which the existence of IS is
expected. However,
the additional degree of freedom, together with the two new
parameters, namely the
distance between the dots and the tunneling strength, may give rise
to new phenomenology.
For instance, Yang
et al. \cite{yan} suggest an experiment to test the quantum coherence
of a special stable two-level system built in a DQD
submitted
to an adjustable interlayer bias voltage, which demonstrates suitable
conditions for serving as quantum computing bits.
Moreover, correlation effects can be
experimentally detected in the far infrared range (FIR) using uniform
electric fields
with non-vanishing component along the vertical direction as the
generalized Kohn theorem, under such condition does not apply
\cite{bar}.

\medskip

This paper is organized as follows. In Section II we describe the
model used in our calculation and analyze the Hamiltonian of the
system. In Section III, after the identification of the
incompressible states of interacting electrons, we begin with a review
of the
results previously obtained for single dots and show next our main
results
for double dots, which cover a wide range of input parameters.
Finally, in Section IV we compare our findings with previous
results in the literature and draw our conclusions.

\medskip

\subsection*{II. The Hamiltonian}

We consider two identical two-dimensional quantum dots (in a vertical
configuration) confined to the XY-plane by equal parabolic
potentials and submitted to a strong magnetic field directed along an
arbitrary direction. The Hamiltonian of the
system reads,

\begin{equation}
H=H_0+H_t+H_{e-e}
\end{equation}
where $H_0$ is the single-particle part which contains the kinetic
contribution, the confining potential and the Zeeman term. We adjust the
input parameters in such a way
that Landau level mixing is negligible. Then, in
second quantization formalism is given by

\begin{equation}
H_0=\,\alpha M\,+\,\beta N\,-\,\Delta_ZS
\end{equation}
where

\begin{equation}
\alpha=\frac{\hbar}{2}(\sqrt{\omega_c^2+4 \omega_0^2}-\omega_c)\,\,\,,
\end{equation}
\begin{equation}
\beta=\frac{\hbar}{2} \sqrt{\omega_c^2+4 \omega_0^2}\,\,\,,
\end{equation}
and
\begin{equation}
\Delta_Z=\mu_B g B
\end{equation}
$\omega_0$ being the confining potential frequency, $\omega_c$ the
cyclotron frequency given by $\omega_c=eB/m^*c$ ($m^*$ is the
effective electron mass, $B$ the magnetic field and $e$ and $c$ the
electron charge and the
speed of light in vacuum respectively), $\mu_B=e\hbar/2mc$ the Bohr
magneton and $g$ the Land\'e factor (we will consider
$\mid g\mid=0.44$ whenever the Zeeman term is included).
$M=\sum_{i=1}^{N} m_i a^+_{\sigma
_i}a_{\sigma_i}$ is the total angular momentum and $N$ is the
total number of electrons. $a^+_{\sigma_i}$ creates a single particle
state and
$\sigma_i$ refers
to the three indexes that characterize the single particle wave
functions: angular momentum, spin and isospin ( $s$ or $a$ associated to
symmetric and antisymmetric combinations of wave functions
concentrated in each dot: right and left). The tunneling term is given
by

\begin{equation}
H_t=-\frac{\Delta_t}{2} X
\end{equation}
where $\Delta_t$ is the energy gap between the symmetric and
antisymmetric states in the noninteracting system and
$X=N_S-N_A$ is given
by the balance between symmetric and antisymmetric states. Finally the
two-body interaction part of the Hamiltonian is given by

\begin{equation}
H_{e-e}=\frac{1}{2}\,\sum_{ijkl}\,\sum_{\Lambda=1}^3
\,\,V_{ijkl}^
{(\Lambda)} \,\,a^+_{\sigma_i} a^+_{\sigma_j} a_{\sigma_l}
a_{\sigma_k}\,\,\,\,,\,\,\,\,\,V^{(\Lambda)}_{ijkl}=\langle ij\mid
V^{(\Lambda)}\mid kl \rangle
\end{equation}
where
the index $\Lambda$ is used to distinguish between the three
different
possibilities: (i) $V^{(1)}=0$ when only one change of a single
particle
isospin takes place, (ii) $V^{(2)}=\frac{1}{2}(V_{rr}+V_{rl})$ when
both isospins
remain unchanged and (iii) $V^{(3)}=\frac{1}{2}(V_{rr}-V_{rl})$ when
both isospins
are changed \cite{bar}. $V_{rr}$ and $V_{rl}$
are the
intra and inter-dot Coulomb potentials respectively, which are given by

\begin{equation}
V_{rr}=\frac{e^2}{\epsilon r}
\end{equation}
and
\begin{equation}
V_{rl}=\frac{e^2}{\epsilon (r^2+d^2)^{1/2}} \quad\quad\quad ,
\end{equation}
$d$ being the distance between the dots along the $z$-direction,
$\vec{r}$ a 2-dimensional vector and $\epsilon$ is the dielectric
constant of the host semiconductor. We have assumed
Dirac- delta distributions along the $z$-direction
and have taken as a basis, Slater
determinants built up from Foch-Darwin single particle wave functions
projected on the LLL \cite{jac}.
The diagonalization can be
performed in separated subspaces characterized by three well defined
quantum numbers: the total angular momentum $M$, the total spin $S$
along the direction of the field $\vec{B}$ and the parity $P$ related
to the reflexion symmetry with respect to the plane midway between the
dots (
$P$ defined as
$P=(-1)^{X/2}$ for even $N$ and $P=(-1)^{(X+1)/2}$ for odd $N$ ). We
will define the set $(M,S,P)$ as a configuration.

\medskip

The eigenstates within each
configuration
are determined by
$H_{e-e}+H_t$ alone, and the role of the constant term given by $H_0$
is
to shift the eigenenergies as a whole without changing their relative
order.

\medskip

\subsection*{III. Incompressible states in the LLL.}

\subsection{Single QD.}

\medskip

Before studying the IS's in DQD's, we briefly
review previous work on single QD's and its
consequences. For a QD an
IS with total energy $E$ and characterized by $(M,S)$ is
identified as the one which has the
following singular property\cite{mak}: the lowest excited state with quantum
numbers $(M+1,S)$ has energy $E+\alpha$. That is to say,
the energetically most favorable way to excite an IS increasing
its total angular momentum by one unit is by moving the system as
a whole, namely by increasing by one unit the angular momentum of the
center of mass (CM) only and leaving the internal structure
unchanged.
This
characteristic
was nicely recognized analyzing the Coulomb contribution to the total
energy of a full polarized QD as a function of $M$. A
periodical arrangement of plateaux
(steplike structure) in the otherwise decreasing curve signaled the
values of the
magic angular momenta \cite{mak}.
Furthermore, the variation of the
magnetic field (or the confining potential) did not drive the
ground state (GS)
through all neighboring values of $M$ but through the
sequence of magic values only
\cite{haw}.

\medskip

This scenario corresponds to the regime
characterized by a
filling factor lower or equal to one, defined as
\cite{mak}

\begin{equation}
\nu=\frac{N(N-1)}{2M}
\label{nu}
\end{equation}
which involves the minimum possible value of the total angular momentum
for a full polarized QD given by $\,\,M_{min}=N(N-1)/2\,\,$ (the
"compact
state") and the angular momentum $M$ of the magic state.
Some care must be taken for low values of $B$ for which the assumption
of the LLL regime is not fulfilled. A suitable way to check this
condition is by making sure that the energy of the highest
single-particle occupied state is much smaller than
$\,\,\omega_+\,=\,\frac{\hbar}{2}(\sqrt{\omega_c^2+4\omega_0^2}+
\omega_c\,)\,\,$,
which is the energy gap
between Landau levels for non-interacting electrons.
The filling factor refers to the number of sublevels
occupied within the LLL. There are two sublevels
(spin up and down) in the case of a single QD and four (two for spin
and two for isospin) in the case of a DQD. In general,
for regimes in which several sublevels are occupied, the filling
factor of a QD is not well defined.
GS's which are
not related with IS are also possible under such
multiple-sublevel occupancy.

\medskip

The sequence of magic filling factors depend on the number of
electrons, for $N=3$ the values of $\nu$ are
$\nu=1,\frac{1}{2},\frac{1}{3},\frac{1}{4},..$
or for $N=4$ they are $\nu=1,\frac{3}{5},\frac{3}{7},\frac{1}{3},..$,
in both cases related to the magic angular momentum given by

\begin{equation}
M=\frac{1}{2}N(N-1)+jN
\label{M}
\end{equation}
where $j$ is a positive integer number. It turns out that the analysis
of the Coulomb contribution to the total energy as a function of $M$
gives exhaustive and precise information about the magic values of
the angular momentum and hence about the magic filling factors. The magic
values of $M$ are the initial values of the plateaux. However, no
information about the total spin of the IS's comes
from the
previous analysis. In the $N=3$ case, for a QD, the sequence of GS's
is always full polarized $(S_z=3/2)$ if the Zeeman term is
included in the Hamiltonian
(with $\mid g \mid =0.44$ ) or in
contrast, oscillations between
$S_z=3/2$ and
$S_z=1/2$ were obtained if no Zeeman term is included in the
calculation \cite{mak,haw}. However, in the last case, the
changes in spin and angular momentum do not appear simultaneously.

\medskip

\subsection{DQD for $d\sim l_B$.}

\medskip

For a DQD we have a richer parameter space to be explored as,
in addition to the parameters of a single QD,
$\Delta_t$ and $d$ also enter the Hamiltonian, which
open new possibilities for IS's to exist.
We will focus on the phase diagram ($\Delta_t\,/\,d$) for
standard values of the remaining input parameters.
Due to the fact that Coulomb interaction and changes in parity are
coupled processes in a DQD, we define the "interaction"
energy as the Coulomb
plus the tunneling contribution (C+T).

\medskip

For $d\,\sim \,l_B$, where $l_B$ is the magnetic length
given by
$\,l_B=\sqrt{\frac{\hbar}{m^*(\omega_c^2+4\omega_0^2)^{1/2}}}$,
the pure Coulomb contribution to the total energy (
$\alpha=\beta=\Delta_t=\Delta_Z=0$)
as a function of $M$, is a decreasing function without plateaux as it is
shown in the curves (a) in Fig.1 and 2 for $N=5$ and $N=6$
respectively
(energies are given in units of $u=e^2/\epsilon
l_B$).
Figs.1A and 2A correspond to parity $P=1$ and Figs.1B and 2B to
$P=\,-1$. All four cases refer to full polarized systems (S=5/2 for
N=5 and S=3 for N=6) . For each value of $M$, the energy displayed is
the lowest within the configuration
$(M,S,P)$.
The absence of plateaux can be understood as follows.
Since $\Delta_t=0$,
the number of electrons in each dot is a well defined number.
Hence, in order to increase the total angular momentum by one unit,
the angular momentum of either dot must be increased by one unit, which
unavoidably increases the typical distance from the electrons of one dot
to the ones of the other dot, and,
therefore, decreases the inter-dot Coulomb energy.

\medskip

According to Figs. 1 and 2, it is
necessary to include a sizeable tunneling contribution in order to
obtain a sequence
of plateaux, which, furthermore, only occur for $P=-1$.
 Indeed, from a series of
calculations for $N=5$ (not shown in Figs.1), which correspond to
a variation of $\Delta_t$ from $0$ to $2\,\, meV$ by small steps, we
see a number of plateaux gradually appearing as $\Delta_t$ increases.
We find
that from $\Delta_t=2\,\, meV$ to $0.8\,\, meV$ the sections
from $M=10$ to $M=11$ and from $M=15$ to $M=16$ are exact
plateaux. For $\Delta_t=0.4\,\,meV$ they are approximately
flat and for $\Delta_t=0.2\,\,meV$ they disappear. However
in all cases the curves are abruptly decreasing before $M=10$ and
between magic values. That is to say, we do not find any extra value
of magic $M$ different from those given by Eq.(\ref{M}).

\medskip

In order to obtain information about the spin and parity of the
IS's, we calculated the Coulomb plus tunneling
contribution for
all possible configurations. Fig.3 (for $N=5$) shows that the
sequence of plateaux appear only when the system is fully polarized in
spin and have parity $P=-1$ (similar results were obtained for $N=6$).
 Furthermore, although the parity $P=-1$
for N=5 can
be obtained from different values of $X$, i.e.,
$X\,=\,N_S\,-\,N_A\,=$ 5, 1 or -3, the
occupancy of the single-particle states for such incompressible
GS turns out to be $X\,=\,5$ only, namely the system
is always
full spin and isospin polarized.
This suggests that
the GS that are IS's will not present variations in
$S$ or $P$ as
$B$ increases.
This last suggestion was confirmed,
for Zeeman contribution different from zero, by an explicit calculation
of the GS vs $B$, which turns out to be always full spin and isospin
polarized. Fig.4 displays the total energy of the GS as a
function of
the magnetic field. The arrows point to the places where the
angular momentum jumps from one magic value to the next one, leaving
the spin and parity unchanged.
In the inset we show $E_{GS}-\beta N$
in order to compare with other publications which omit the
$N$-dependent term.
The nearly
monotonous function of $B$ is due to the fact that,
in the absence of spin or isospin transitions, the interaction  energy
has a negligible influence in the plot and hence the
evolution of the system is driven by the monotonous increasing term
$\beta N$ which is much more important than the decreasing term
$\alpha M$, which would produce kinks at the transition points, as it
is shown
in the inset ($E_{GS}-\beta N$ vs $B$). In brief, the
full spin
and isospin polarization appears to be a well defined attribute of
these IS's that result from many-body effects.

\medskip

As shown in Fig. 3A the interaction energy appears to be
degenerated at the magic values with respect to the three possible
spin polarizations. Since the curves that belong to $S=\frac{3}{2}$
and $\frac{1}{2}$ have lower energy at the end of the plateaux, the
final balance of energy depends critically on the relation between this
difference of interaction energy, the kinetic and the Zeeman terms.
That is to say, an IS that is the GS for a given
value of $B$ and $\omega_0$ will remain as GS as $B$ increases or
$\omega_0$ decreases only if
\begin{equation}
E_{C+T}(M,\frac{3}{2},1)-E_{C+T}(M+1,\frac{3}{2},1)\,\,<\,\,
\alpha\,\,+\,\,\Delta_Z
\end{equation}
or otherwise the new GS will be a compressible not fully spin
polarized state at $M+1$. Hence, as $B$ or $\omega_0$
change, the GS can be driven into compressible zones in contrast with
the results obtained for single QD's.

\medskip

The single particle occupancies of the $m$-values for the
first three IS's for $N=5$, calculated at $B$ = 4, 7 and 9 T
respectively are shown in Fig.5. The first GS for $M=10$ ($\nu=1$) is
the compact full polarized state which belongs to a one dimensional
subspace and, as a consequence, no correlation is involved as one
Slater determinant produces the exact solution. Moreover, the density
is a "dome" shape circular symmetric distribution  without any
structure. As $B$ grows, the angular momentum changes from $M$ to
$M+N$, all the electrons jump together moving away from the origin
and forming a ring. The dimension of the GS subspace increases and
as the relative weights of the different Slater determinants within
the expansion of the GS become significant for different elements of
the bases, namely, the correlation becomes important.

\medskip

\subsection{DQD: Phase diagram $\Delta_t/d$.}

\medskip

So far we have explored the situation $d\,\sim\, l_B$. In order to get
the complete scenario of IS's in a DQD, we have also investigated in
detail the remaining regions of the phase diagram ($\Delta_t\,/\,d$)

\medskip

In Figs.6-9 we follow, for $N=3$, the variation of the interaction
energy vs $M$ ($E(C+T)/M$) as $d$ and $\Delta_t$ change.

\medskip

From A to B (Fig.6), as it was just discussed for the case $d\sim
l_B$, the plateaux emerge as $\Delta_t$ grows from zero until they
are well defined at $\Delta_t=0.11\,\,u$, that is to say, for values
of $M$ given by Eq.(\ref{M}) the system evolves from
compressible to incompressible states.
For $d$ close to zero (at the left of point $B$) and $\Delta_t$ large,
all the electrons are in the symmetric state. As a consequence, the
interaction energy
of a single QD can be reproduced with high accuracy by the addition
of
the constant contribution $\Delta_t\,X/2$ to the energy of the DQD in
this region.

\medskip

From A to D (Fig.7), tunneling between the two dots is not
allowed. Starting from a curve without plateaux for small
distances ($d\sim 10 \AA$), we move across the transition
regime with a gradual formation of new plateaux at
$M=M_R+M_L$  where $M_R$ and $M_L$ are the magic numbers of
single QD's. We come close to the point D at $d=1000 \AA$
which show the features of two decoupled dots with $N=1$ (with no
contribution to the Coulomb term) and $N=2$ (with magic
numbers $M=1,3,5,7..$) respectively. This regime in which
tunneling is forbidden has been previously studied for
double layers and special attention has been devoted to the $\nu=1$
case \cite{cha1,mur,yan}. For a double layer
the incompressible state $\nu=1$ is observed for values of $d$
about the magnetic length \cite{mur}. Furthermore, as $d$
increases, the state exhibits a phase transition to a compressible
one. The difference between the two cases comes from the fact
that,
as it was previously discussed, when $\Delta_t\sim 0$ $M_R$ and $M_L$
are
well defined quantum numbers, the increase by one unit of $M$ means
the increase of $M_R$ or $M_L$ (but not both) changing, in a DQD, the
relative position of the electrons in each dot and so decreasing the
Coulomb interaction which prevents the formation of a plateau.
However, this is not the case for a double layer in which the shift of
charge due to the change of angular momentum  does not change the
relative
inter-layer distribution of charge, allowing for the appearance of
plateaus.
For large $d$ the two layers decouple and hence one would expect (for
total $\nu=1$) two $\nu=1/2$ IS's. However, since fractional QHE of
$\nu=1/2$ is not observable for a single layer, these states were not
identified in Ref.(\cite{mur}).

\medskip

From B to C (Fig.8), although the distance $d$ grows, the
sequence of magic numbers typical of a DQD does not disappear
due to the relative large value of $\Delta_t$
($\Delta_t=0.229\,\,u$). An exceptional case appears for
$d\,\,> 500\AA$ at $M=1$ (for $N=5$ the analog exceptional case
appears at $M=4$). It turns out to be the only IS
within the LLL regime which does not fulfill the general rule of
being full spin and isospin polarized. The subspace associated to the
appropriate configuration, i.e., ($M=1$, $S=3/2$, $P=1$) is one
dimensional and the only Slater determinant in the bases has one
electron in the symmetric state and two electrons in the antisymmetric
state (or $X=-1$). It is the only IS with no
single dot analog. Our interpretation of the fact that the $M=1$ magic
value appears only for relative large distances is as follows: for
large tunneling and small $d$, the system is closer to a single dot
with $N=3$ than to a DQD of the same number of electrons (as
discussed before). Thus
$M=1$ can
only appear when the Coulomb inter-dot interaction weakens related to
$\Delta_t$ and DQD properties different from those of a single QD
may arise.
Notice however, that for the $M=1$
state to be a GS such a low magnetic
field (or large confining potential) is required that the LLL regime
assumption would not apply anymore. Finally,
even for values of $d$ as large as
$1500\,\,\AA$ (being $\l_B=65\AA$),
we did not find the
transition from
DQD to two decoupled single dots.

\medskip

From D to C (Fig.9), the tunneling increases and
the system of two decoupled QD's with a period of two
typical of the $N=2$ single dot evolves into a DQD
reproducing the period of three typical of a $N=3$ DQD.
During the transition, there is a narrow interval of values of
$\Delta_t$ for which the $E(C+T)$ vs $M$ curve has no plateaux (except
for the $M=1$ case). That is to say, an initially incompressible
GS
would evolve into a compressible state and
again into a IS as $\Delta_t$ increases. This
evolution
takes place as the system changes from two decoupled single dots to a
DQD.

\medskip

Compressible regions have been obtained before by Rontani et al.
\cite{ron} for DQD with finite width. They consider the evolution
of the GS of the system of $N=6$ as $d$ increases for $\Delta_t$
exponentially decreasing with $d$,
which is equivalent to the evolution along a
trajectory from B to D in our phase diagram. They obtain a
small zone of compressibility in the middle, related to the transition
from a regime where the system behaves as a unique coherent system to
a regime of well separated QD's. We observe the same qualitative
behavior along the $B$-$C$ trajectory (which is different to theirs)
although we do not obtain the same magic values.

\medskip

The transition from DQD to two decoupled single QD's as $\Delta_t$
decreases has been observed before by Peeters et al \cite{par1} by
means of a current spin density functional calculation.

\medskip

We have also studied the D to C evolution for N=5 which shows the same
qualitative behavior. As D is approached the structure of plateaux
can be
understood in terms of the IS of two decoupled $N=2$ and $N=3$ single
QD's, although the analysis is much more intricate than for the N=3
case.

\medskip

\subsection*{IV. Discussion}

We have investigated in detail the existence of IS's that result
from the Coulomb many-body effects in a DQD for the entire phase
diagram ($\Delta_t\,/\,d$).

\medskip

An important point in our analysis is the criterium used to identify
IS's. First we want to emphasize that in contrast to the
case of a single layer for which the integer QHE is associated to gaps
of single-particle origin and the fractional QHE is associated to gaps
involving many-body effects, for double layered systems (and thus in
accordance for DQD), single-particle
as well as many-body regimes can be related to QHE at the same filling
factor by the tuning of appropriate sample parameters \cite{mur}. The
IS's we are interested in are those associated with
$e$-$e$ interaction
(coupled with tunneling) and thus signaled somehow in the variation
of
the interaction energy with $M$. As discussed in the previous section,
we define the interaction
energy as the Coulomb plus tunneling contributions, and require IS's
to preserve the interaction energy when the angular momentum $M$ is
increased by one unit.
We want to stress that this is not equivalent to identifying magic
$M$ from
the kinks of the lowest energies of each
configuration as a function of $M$ or from the kinks of the
variation of the absolute GS energy as a function of $B$,
as it has been used in the literature \cite{ima1,ima2} to identify
correlated IS's.

\medskip

Our criterium is equivalent to the one used by
Laughlin in
Ref.(\cite{lau}) for the single layer, as we discuss below.
The Hamiltonian is separable into the CM
and the relative coordinates and as a consequence, the total
angular momentum can be analyzed as $M=M_{CM}+M_{rel}$ and the energy
as $E_{tot}=E_{CM}+E_{internal}$.
For three two dimensional electrons, Laughlin
obtains that the {\it internal} energy as a function of the {\it
relative} angular momentum has downward cusps at special (magic)
values ($M=3,6,9,12,..$).
These magic values appear to be
related to incompressibility: the area of the system defined as the
area of the triangle determined by the correlated positions of the
electrons within these states changes discontinuously as pressure is
applied. At the downward cusps,
$E_{internal}(M_{rel})\,\,<\,\, E_{internal}(M_{rel}+1)$
for the lowest energy states of each configuration. They are the
only states for which the increase of $M_{rel}$ by one unit requires a
positive amount of internal energy.
In order to show that our criterium \cite{mak} is equivalent to Laughlin's \cite{lau}, notice first that
 $E_{C+T}(M)$ only depends on $M_{rel}$ and $M_{rel}\leq M$ .
 Since
$E_{C+T}(M)$ is defined as the minimal energy among those
of the states with total angular momentum $M$, it implies it is the
minimal energy among all states with relative angular momentum
$M_{rel}\leq M$.
Hence, given $M$ and $E_{C+T}(M-1)$,  $E_{C+T}(M-1)
\not= E_{C+T}(M)$ implies $E_{C+T}(M-1) > E_{C+T}
(M)$ and furthermore $M=M_{rel}$. Since $E_{C+T}$ reduces to
$E_{internal}$ for a
single layer,  if $E_{C+T}(M-1) > E_
{C+T}(M)$ then  $E_{internal}(M_{rel}-1) > E_{internal}(M_{rel})$
 for $M_{rel}=M$. Namely, negative slopes in our plots imply negative
slopes in Laughlin's. If, on the contrary, $E_{C+T}(M-1) = E_
{C+T}(M)$, then $M_{r
el}$ has not changed and $M$ contains at least one unit of CM angular
momentum. Being $E_{C+T}(M)$ the minimal energy with total
angular momentum $M$, it implies that
any state with $M_{rel}=M$ has larger energy than $E_{C+T}(M-1)$.
If, in addition, $E_{C+T}(M-2) > E_{C+T}(M-1)$, as it is
always the case in our plots, then $M-1$ contains only relative angular
momentum, as shown above. Then, for the single layer, $E_{C+T}(M-1)
= E_{C+T}(M)$ implies $E_{internal}(M_{rel}-1) < E_{internal}(M_
{rel})$ for $M_{rel}=M$.
Namely plateaux in our plots imply positive slopes in Laughlin's, which concludes our proof.
On the other hand, there is no guarantee that downward cusps in
the curve $E_{tot}$ vs $M$ are related to plateaux in the curve
$E_{C+T}$ vs $M$. It is enough to have $\alpha$ $>$
$E_{C+T}(M)\,\,-\,\,E_{C+T}(M+1)$ at a non-magic $M$
to obtain there a downward peak in $E_{tot}$ vs $M$ which is not
related to a plateau in $E_{C+T}$ vs $M$.

\medskip

For $d\,\sim\,l_B$ the fact that the magic $M$ follow Eq.(\ref{M})
is in conflict with the claim made in Ref. \cite{ima1,ima2} that
extra magic values for $M$
(depending on the value on the
tunneling strength) exist in this regime. The authors of
Ref.(\cite{ima1,ima2})
identified IS's with downward
cusps of the total energy as a function of $M$, i.e. the interaction
energy (our curve) with the addition
of the single-particle contribution. In Fig.10 we show the two
possibilities for $N=5$. It is clear
from the upper curve that some
downward cusps, which would be identified as IS's by the criterium of
Ref.\cite{ima1,ima2}, do not actually correspond to IS's in our
criterium.

\medskip

In order to make sure that the discrepancies with Ref.
\cite{ima1,ima2} are only due to the different criteria to identify
IS's, we have reproduced their results (see Fig.11).
To be more precise, we performed the calculation for $N=3$ and the
same input parameters as those used in Ref.\cite{ima2}
( $N=3$, $B=15\,T$,
$d=200\,\,\AA$, $\hbar \omega_0=3\,meV$
and
$\Delta_t=0.2\,\,meV$ ).
In Fig.11A the interaction energy contribution versus $M$ is shown.
Due to the
low value of the tunneling contribution, the plateaux that will
appear, for larger values of $\Delta_t$, at $M$= 3, 6 and 9 are still
not
visible and the only ones that already appear are $M$= 12 and 15. If
the
kinetic contribution is added, it comes out that the GS is
at $M$= 5 (see Fig.11B) at the lowest downward cusp of the total
energy
in accordance with Ref.\cite{ima2}.

\medskip

A word of caution should be given here as a number of relevant papers
exists in the literature \cite{ima2,seki,yann,mak2} which use the term
{\it magic angular momentum} to denote the angular momentum $M$ which
displays downward cusps in the curve $E_{tot} (M)$.
The corresponding states enjoy enhance stability and have been the
subject of intensive studies. In particular Refs. (\cite{seki,yann})
provide characterizations
of these states ranging from small values of $M$, where the
fractional
quantum Hall  regime is sometimes identified, to large values of $M$,
where strip-like structures and Wigner molecules
seem to appear (see Ref. (\cite{mak2}) for a review and Ref. (\cite{shi})
for related work on layers). However, only a subset of these states
fulfills our criterium of incompressibility and the magic $M$ displayed
in (\ref{M}) correspond to this subset only.

\medskip

Our conclusions  can be summarized as follows:

\medskip

(i) The downward cusps obtained by Laughlin in Ref.(\cite{lau}) turn
out to be equivalent to the plateaux of the curve $E_{C+T}$ vs $M$.

\medskip

(ii) All the incompressible states are full spin and isospin polarized
(except the $M=1$ case for $N=3$).
Since a single QD full spin polarized and a DQD full
spin ans isospin polarized are systems with no extra
degrees of freedom aside from angular momentum, we expect a similar
behavior for the electronic distance quantization as that obtained
in Ref(\cite{lau}).

\medskip

(iii) An exceptional incompressible state was found for $d/l_B\geq 8$
at $M=1$ and $X=-1$ for $N=3$ and $M=4$ and $X=1$ for $N=5$. This is
the only one that is not full isospin polarized and does not have a
single QD analog. However, for it to be a GS  values of the
input parameters that do not fulfill the assumption of the LLL regime
are required.

\medskip

(iv) For $d\sim l_B$, it is not possible to obtain IS's if the
tunneling is small.

\medskip

(v) For small $\Delta_t$, as $d$ grows, the DQD evolves into two
decoupled single QD's. New magic values appear which correspond to the
addition of magic numbers of two decoupled QD's.

\medskip

(vi) An IS at $M$ will remain as the GS under changes of $B$ or/and
$\omega_0$ if the condition
\begin{equation}
E_{C+T}(M)-E_{C+T}(M+1)\,\,<\,\,\alpha+\Delta_Z
\end{equation}
is fulfilled. In general, however, the variation of $B$, $\omega_0$ or
$d$ can drive the GS from incompressible into compressible zones
of the
phase diagram. This behavior differs from the well known properties
of a single QD for which the variation of the GS as $B$ or $\omega_0$
change is driven through incompressible states only, skiping all
non-magic values of $M$.

\medskip

(vii) Whenever the GS is full spin and isospin polarized, it is an IS.
In other words, the IS's are the only possible GS for $\nu$ lower
that one. However for $\nu\,>\,1$ other GS are possible.

\medskip

Before closing, let us briefly elaborate on the last point. Notice
that the following situations are also possible:
(a) GS's with $M$ that fulfills Eq. (\ref{M}) and are not
IS's due to the fact that from $M$ to $M+1$ there is not a
plateau. This condition can be obtained for very low values of
$\Delta_t$, for example, for $N=5$, $M=10$, $B=5\,T$,
$\hbar \omega_0=2.6\,meV$,
$\Delta_t=0.2\,\,meV$ and $d=20\,\AA$.
(b) Ground

states with
$M$ not given by Eq. (\ref{M}) for which the system is not full spin or
isospin polarized,
(or $\nu\,\,>\,\,1$ ) and $E(C+T)$ has not a constant evolution from
$M$ to $M+1$. This is the case for example for
$N=5$, $M=13$ ( $B=6T$,
$\hbar \omega_0=2.6\,meV$ $d=20\AA$ and
$\Delta_t=5.86\,\cdot 10^{-3}\,meV$). In the last case,
$S=N/2$ and $P=-1$,
however
$X\neq N$.

\medskip

Notice also that the previously discussed states are not the only
possible GS's within the LLL regime. For instance,
if the confining
potential is strong enough, other types of GS's are
possible
like the ferromagnetic, canted and symmetric states (all of them
with $\nu=2$) first studied
in double layers \cite{das} and latter recognized in DQD \cite{tej}.

\medskip

Finally, let us note that the correlation plays an increasingly
important role as the magnetic field
grows up.
The interaction energy
and correlation effects can be experimentally tested by
uniform
electric fields with non-vanishing component along the z-direction due
to the fact that under this condition, the Kohn theorem does not
apply \cite{bar} and the FIR spectroscopy becomes
sensitive to the internal structure.

\vskip6mm

We gratefully acknowledge C. Tejedor and L. Mart\'\i n-Moreno
for the code used for the Hamiltonian diagonalization.
This work has been performed under Grants No. BFM2002-01868 from DGESIC (Spain),
 No. FPA2001-3598 from MCyT and Feder (Spain)
, and No. 2001GR-0064 and No. 2001SGR-00065
from Generalitat de Catalunya.

\eject

\eject

\begin{figure}[htbp]
 \begin{center}
  \psfig{figure=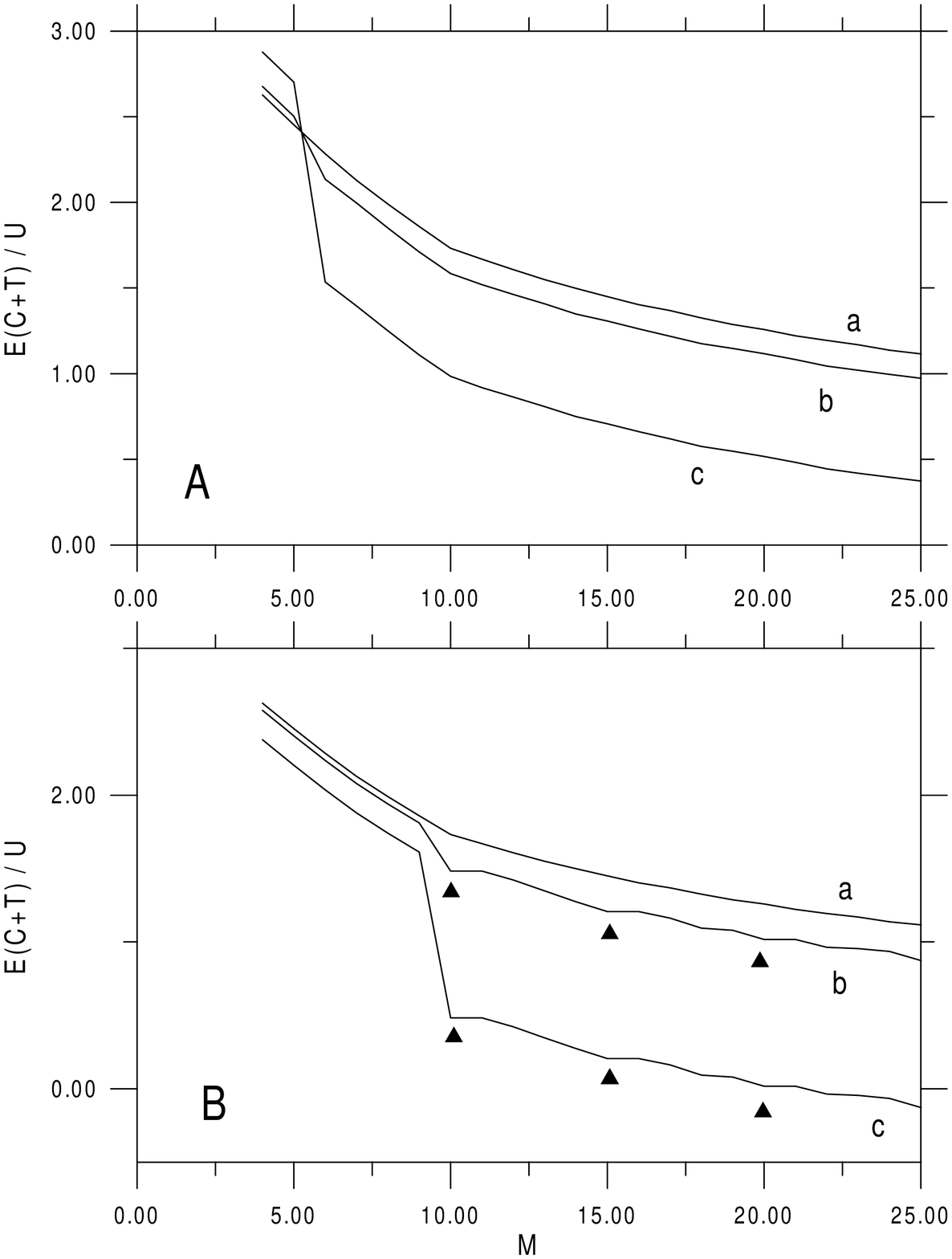,width=12.5cm}
 \end{center}
\caption{A: Coulomb plus tunneling contribution to the total energy
as a funtion of $M$ for $N=5$, $S=N/2$ and parity $P=1$, for several
values
of the tunneling gap: (a) $\Delta_t=0$, (b) $\Delta_t=2.2\,meV$ and (c)
$\Delta_t=11\,meV$.
B: The same as A for $P= -1$. The triangles
point to the begining of the plateaux.
We have taken
$B=5T$,
$d=20\AA$, and
$\hbar \omega_0=2.6\,meV.$}
\end{figure}
\eject

\begin{figure}[htbp]
 \begin{center}
  \psfig{figure=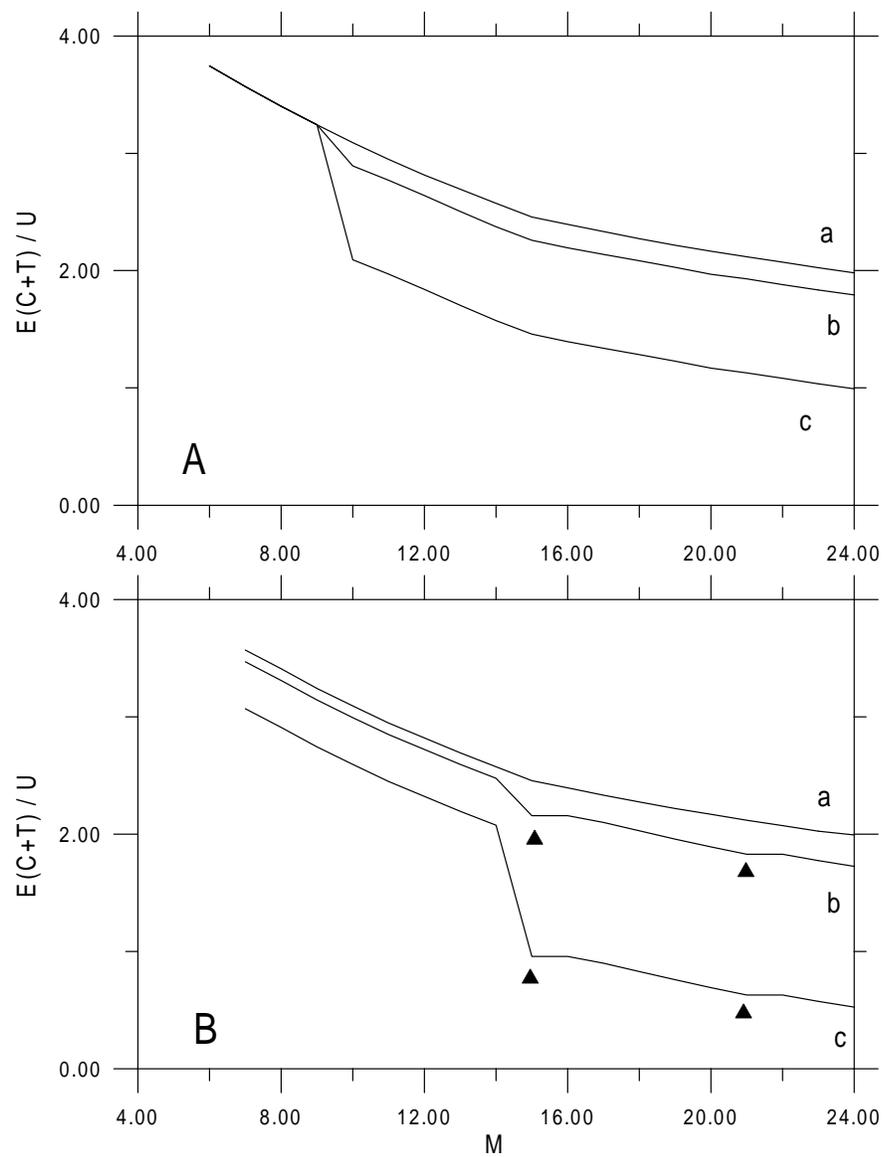,width=12.5cm}
 \end{center}
\caption{The same as Fig.1 for $N=6$}
\end{figure}
\eject

\begin{figure}[htbp]
 \begin{center}
  \psfig{figure=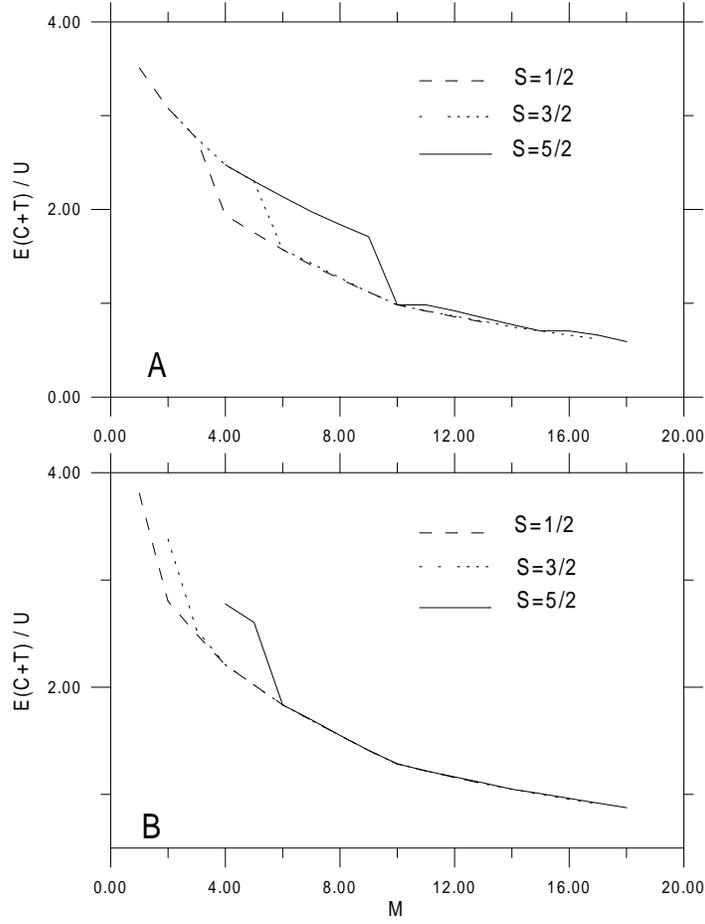,width=12.5cm}
 \end{center}
\caption{A: Coulomb plus tunneling contribution to the total energy
as a function of $M$ for $N=5$ and parity
$P= -1$ for all the possible values of the spin $S$. B: The same as A
for $P=1$. We have taken
$B=5T$,
$d=20\AA$, $\hbar\omega_0=2.6\,meV$ and $\Delta_t=2\,meV$.}
\end{figure}
\eject

\begin{figure}[htbp]
 \begin{center}
  \psfig{figure=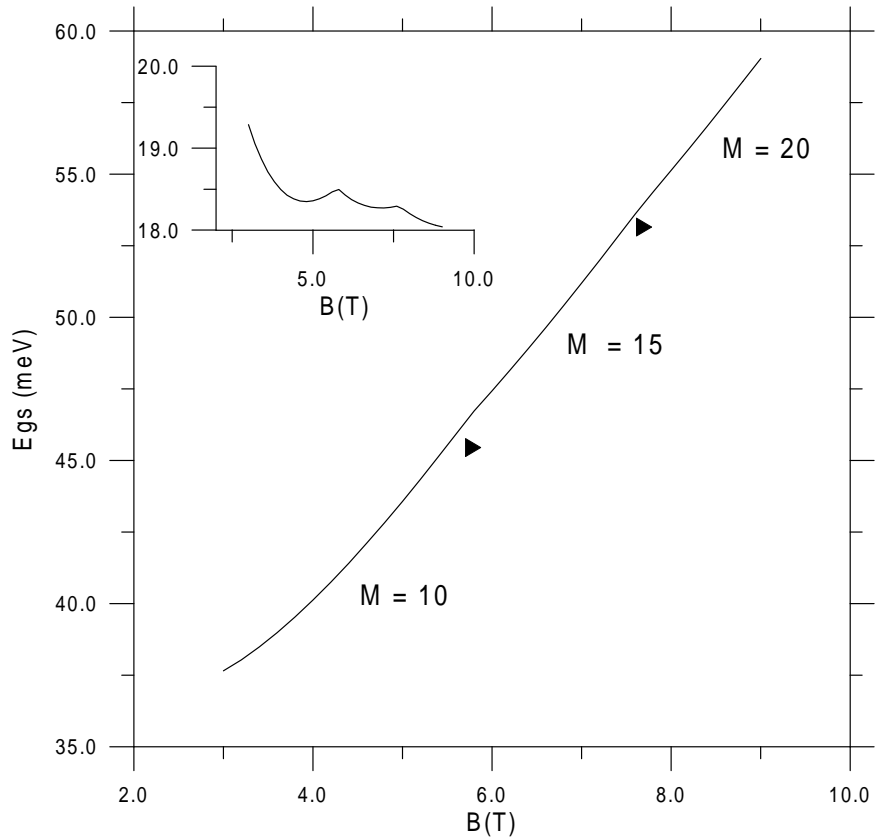,width=12.5cm}
 \end{center}
\caption{A: Ground state energy versus $B$ for $N=5$. The
triangles point to the transitions between magic angular momenta. We use
the same values for $d$, $\omega_0$ and $\Delta_t$ as in Fig.3. Inset:
$E_{GS}-\beta N$ as a function of $B$.}
\end{figure}
\eject

\begin{figure}[htbp]
 \begin{center}
  \psfig{figure=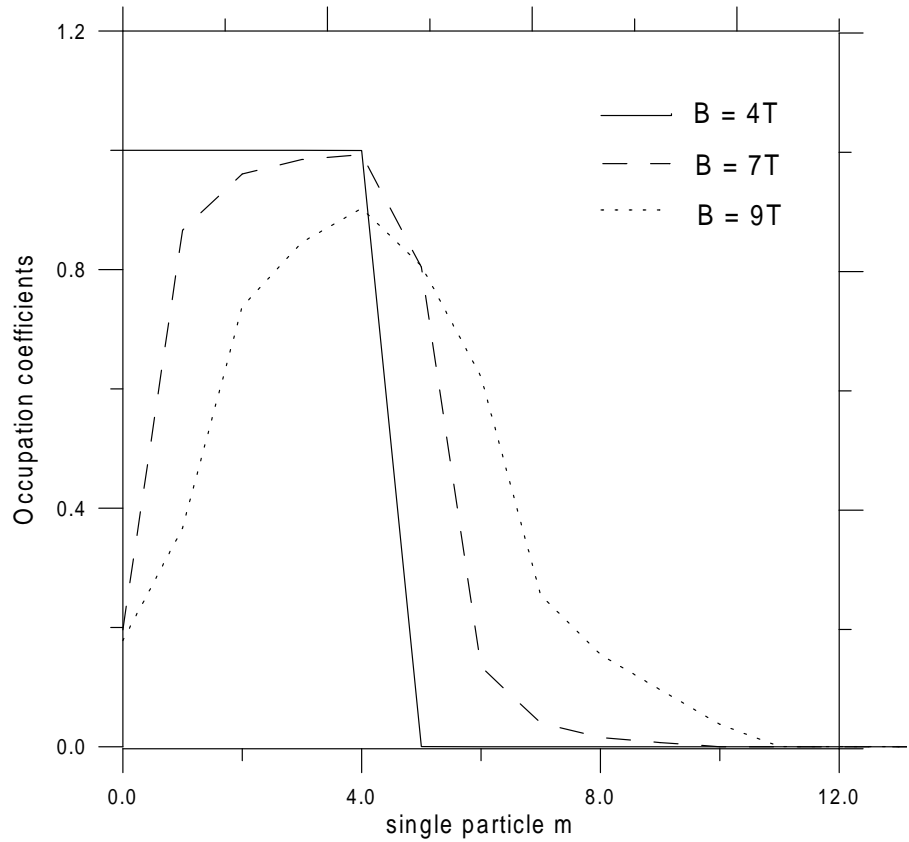,width=12.5cm}
 \end{center}
\caption{Total occupancy of the single-particle angular momentum
states $m$ for $N=5$. The values of $B$ considered correspond to the magic
values $M$ = $10$, $15$ and $20$ respectively. We have taken
$d=20\AA\,$,
$\hbar\omega_0=2.6\,meV\,$ and $\Delta_t=6\,meV$.}
\end{figure}
\eject

\begin{figure}[htbp]
 \begin{center}
  \psfig{figure=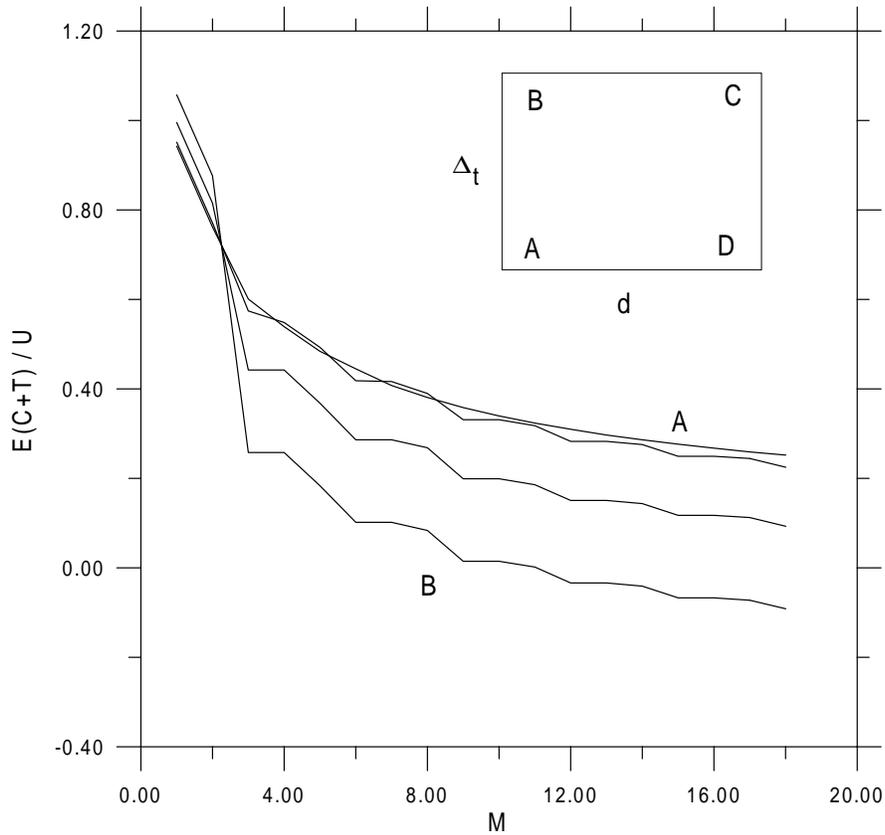,width=12.5cm}
 \end{center}
\caption{ Coulomb plus tunneling contribution
versus $M$ for $N=3$.
From A to B
the tunneling gap is:
$0.0$, $0.018$, $0.106$ and $0.229 u$ respectively. Inset: phase diagram
used.
We have taken $\hbar
\omega_0=2.6\,meV$, $B=15$T, $d=10\AA$, $S=3/2$ and $P=1$
($l_B=65\AA$).}
\end{figure}
\eject

\begin{figure}[htbp]
 \begin{center}
  \psfig{figure=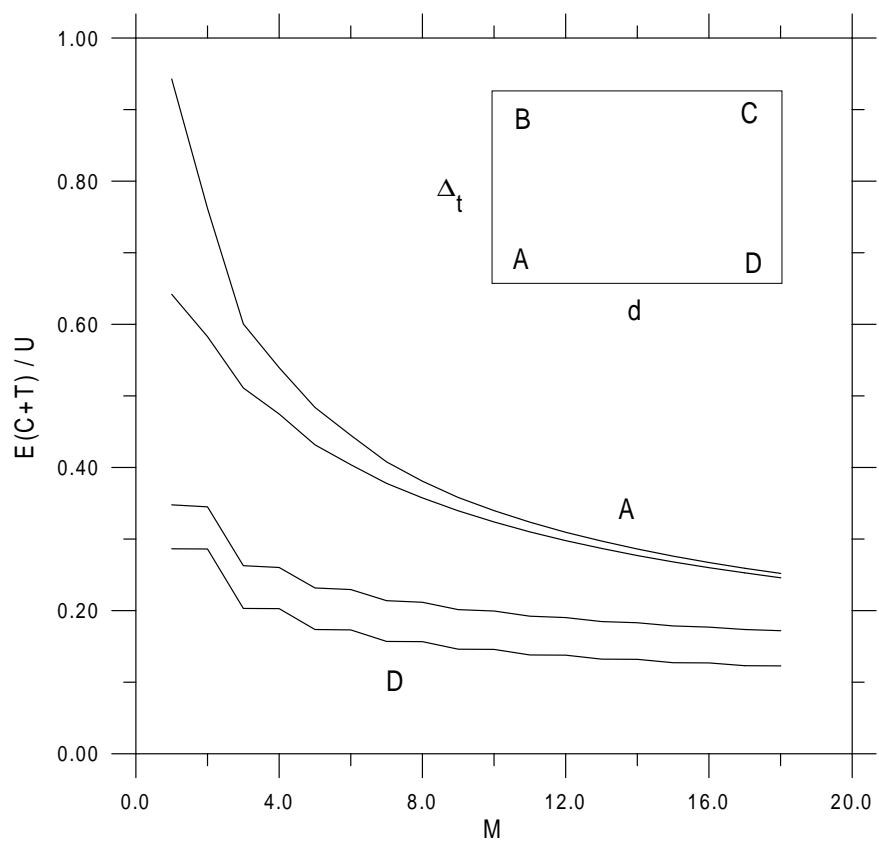,width=12.5cm}
 \end{center}
\caption{ The same as Fig.6 for $\Delta_t=0$.
From A to D the inter-dot distance is: $10$,
$50$, $500$ and $1000$ $\AA$
respectively.  Inset: phase diagram
used.}
\end{figure}
\eject

\begin{figure}[htbp]
 \begin{center}
  \psfig{figure=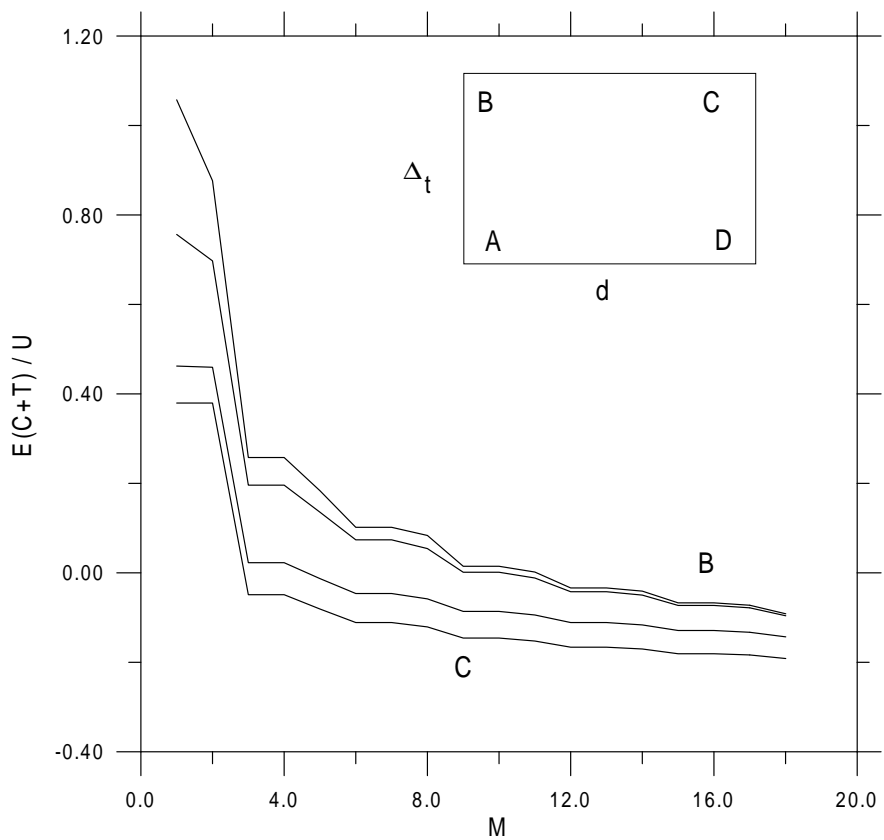,width=12.5cm}
 \end{center}
\caption{ The same as Fig.6 for $\Delta_t=0.229$
$u$. From B to C the inter-dot distance is: $50$, $100$, $500$ and
$1000$ $\AA$ respectively.  Inset: phase diagram
used.}
\end{figure}
\eject

\begin{figure}[htbp]
 \begin{center}
  \psfig{figure=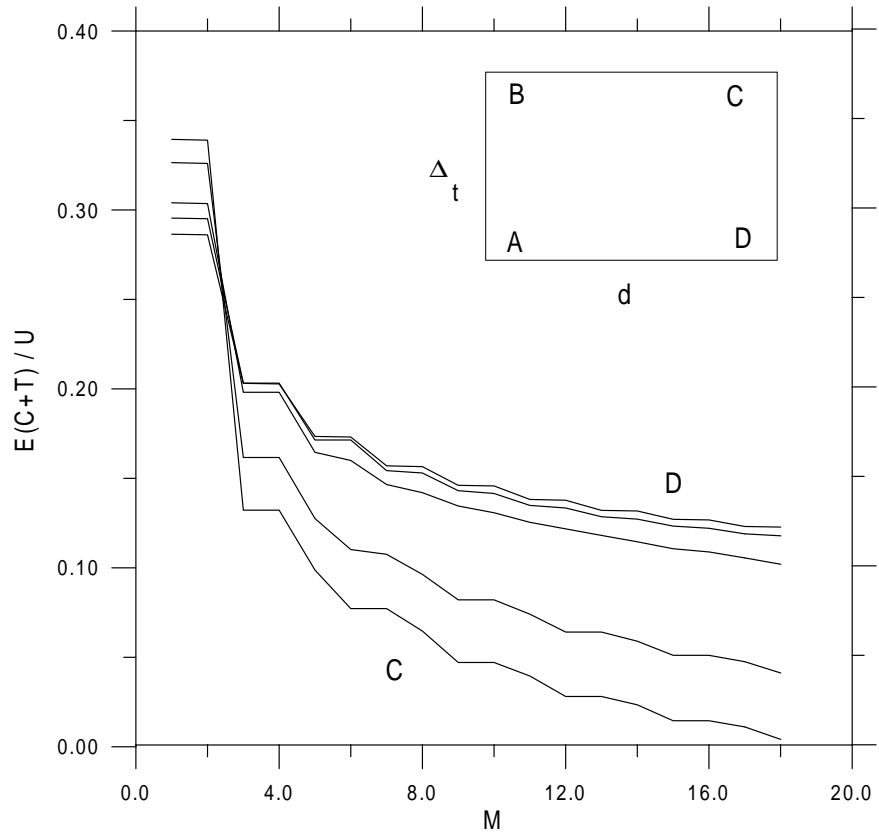,width=12.5cm}
 \end{center}
\caption{ The same as Fig.6 for $d=1000$ $\AA$.
From D to C the tunneling gap is: $0.0$, $0.018$, $0.035$, $0.080$ and
$0.106$
$u$ respectively.  Inset: phase diagram
used.}
\end{figure}
\eject

\begin{figure}[htbp]
 \begin{center}
  \psfig{figure=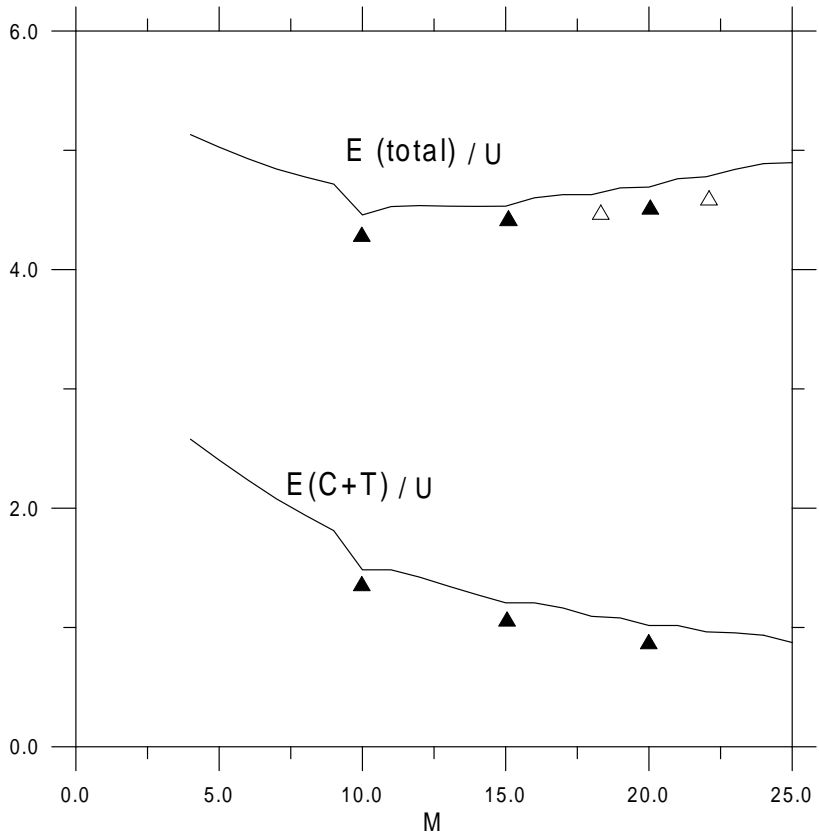,width=12.5cm}
 \end{center}
\caption{Coulomb plus tunneling contribution
(lower curve)  and total energy (upper curve) as a
function of $M$ for $N=5$. The black triangles point to the actual magic values $M$ whereas the white triangles
point to cusps which could be mistaken by them.
The values of $B$,
$\omega_0$, $d$ and $\Delta_t$ are the same as in Fig.3}
\end{figure}
\eject

\begin{figure}[htbp]
 \begin{center}
  \psfig{figure=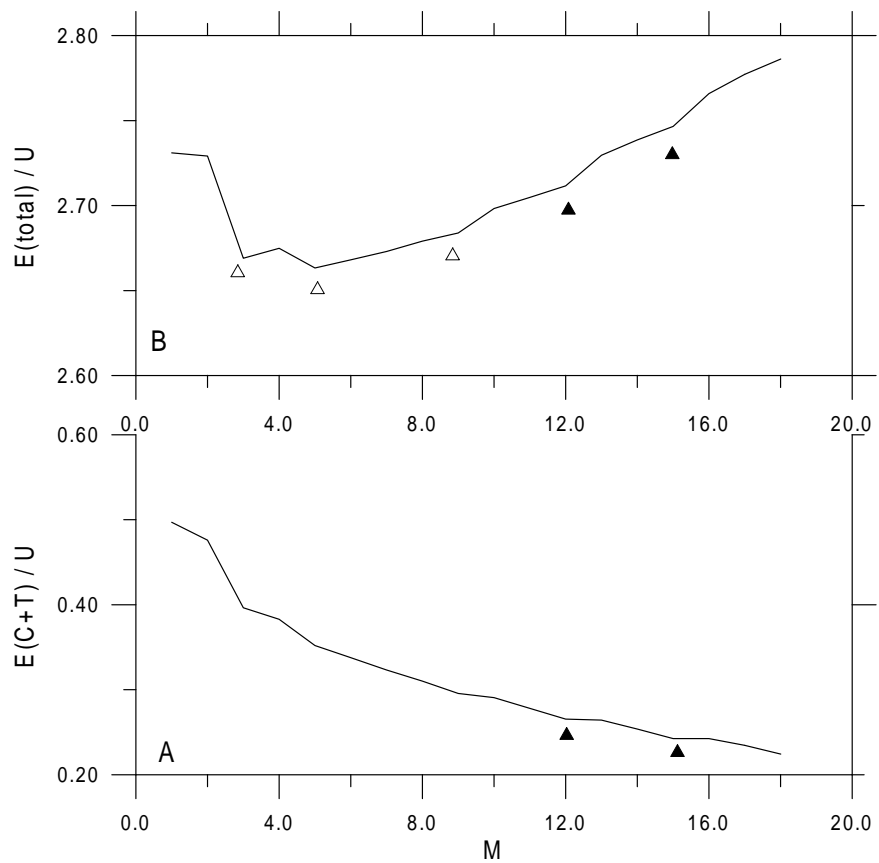,width=12.5cm}
 \end{center}
\caption{A: Coulomb plus tunneling contribution
versus $M$ for $N=3$. B: The same as in A for the total energy.
 The input parameters are given in the text. The black
triangles point to the actual magic values $M$ whereas the white triangles
point to cusps which could be mistaken by them.}
\end{figure}
\eject

\end{document}